\documentclass[aps,twocolumn,,eqsecnum,prb]{revtex4}
\usepackage[dvips]{graphicx}
\usepackage{amsmath,amsbsy,amssymb,graphics}
\usepackage[mathscr]{euscript}

\def\beginABC{\begin{subequations}}
\def\endABC{\end{subequations}}
\let\mathit=\mathscr
\let\mathbf=\boldsymbol

\begin{document}

\title{Anomalous Hall Resistance in Bilayer Quantum Hall Systems}
\author{Z.F. Ezawa$^{1}$, S. Suzuki$^{1}$ and G. Tsitsishvili$^{2}$}
\affiliation{{}$^1$Department of Physics, Tohoku University, Sendai, 980-8578 Japan \\
{}$^2$Department of Theoretical Physics, A. Razmadze Mathematical Institute,
Tbilisi, 380093 Georgia}
%\date{\today}

\begin{abstract}
We present a microscopic theory of the Hall current in the bilayer quantum
Hall system on the basis of noncommutative geometry. By analyzing the
Heisenberg equation of motion and the continuity equation of charge, we
demonstrate the emergence of the phase current in a system where the
interlayer phase coherence develops spontaneously. The phase current
arranges itself to minimize the total energy of the system, as induces
certain anomalous behaviors in the Hall current in the counterflow geometry
and also in the drag experiment. They explain the recent experimental data
for anomalous Hall resistances due to Kellogg et al. [M. Kellogg, I.B.
Spielman, J.P. Eisenstein, L.N. Pfeiffer and K.W. West, Phys. Rev. Lett. 
\textbf{88} (2002) 126804; M. Kellogg, J.P. Eisenstein, L.N. Pfeiffer and
K.W. West, Phys. Rev. Lett. \textbf{93} (2004) 036801] and Tutuc et al. [E.
Tutuc, M. Shayegan and D.A. Huse, Phys. Rev. Lett. \textbf{93} (2004)
036802] at $\nu =1$.
\end{abstract}

\maketitle

\section{Introduction}

The emergence of the Hall plateau together with the vanishing longitudinal
resistance has been considered to be the unique signal of the quantum Hall
(QH) effect\cite{EzawaBook,BookDasSarma}. This is certainly the case in the
monolayer QH system. However, recent experiments\cite{Kellogg04,Tutuc04}
have revealed an anomalous behavior of the Hall resistance in a counterflow
geometry in the bilayer QH system that both the longitudinal and Hall
resistances vanish at the total bilayer filling factor $\nu =1$. Another
anomalous Hall resistance has been reported in a drag experiment\cite%
{Kellogg02}. Though a suggestion\cite{Eisenstein04} has been made that the
anomalous phenomenon would occur owing to excitonic excitations
(electron-hole pairs in opposite layers) in the counterflow transport, there
exists no theory demonstrating these phenomena explicitly in a unified way.

The aim of this paper is to present a microscopic theory of Hall currents to
understand the mechanism of the anomalous Hall resistance\cite%
{Kellogg04,Tutuc04,Kellogg02} discovered experimentally. In the ordinary
theory the current is defined as the N\"{o}ther current, which arises from
the kinetic Hamiltonian. However, this is quite a nontrivial problem in the
QH system\cite{Girvin86B} since the kinetic Hamiltonian is quenched within
each Landau level\cite{Girvin84B,Girvin85L}. There is a Lagrangian approach%
\cite{Stone93IJMPB} to this problem, but it seems quite difficult to go
beyond the one-body formalism within this approach. We propose a new
formalism to elucidate the current in the QH system.

We start with a microscopic Hamiltonian\cite{Ezawa05D} describing electrons
in the lowest Landau level. The intriguing feature is that the dynamics is
determined not by the kinetic Hamiltonian but by noncommutative geometry.
The noncommutative geometry\cite{BookConnes} means that the guiding center $%
\mathbf{X}=(X,Y)$ is subject to the noncommutative relation, $[X,Y]=-i\ell
_{B}^{2}$, with $\ell _{B}$ the magnetic length. We derive the formula for
the electric current from the Heisenberg equation of motion and the
continuity equation of charge based on the noncommutative relation. It
agrees with the standard formula for the Hall current in the monolayer
system. \ 

There arises new phenomena associated with the interlayer phase coherence in
the bilayer QH system\cite{EzawaBook,BookDasSarma}. The bilayer system has
the pseudospin degree of freedom, where the electron in the front (back)
layer is assigned to carry the up (down) pseudospin. Provided the layer
separation $d$ is reasonably small, the interlayer phase coherence\cite%
{Ezawa92IJMPB,Moon95B} emerges due to the Coulomb exchange interaction. The
system is called the pseudospin QH ferromagnet. This is clearly seen by
examining the coherence length $\xi _{\vartheta }$ of the interlayer phase
field $\vartheta (\mathbf{x})$, which is calculated as%
\begin{equation}
\xi _{\vartheta }=2\ell _{B}\sqrt{{\frac{\pi J_{s}^{d}}{\Delta _{\text{SAS}}}%
}},  \label{CoherPpinTwo}
\end{equation}%
where $J_{s}^{d}$ is the pseudospin stiffness and $\Delta _{\text{SAS}}$ is
the tunneling gap. It is observed that the interlayer phase coherence
develops well for $J_{s}^{d}\gg \Delta _{\text{SAS}}$.

It has been argued in an effective theory\cite{EzawaBook} that the phase
current, $\propto \partial _{i}\vartheta (\mathbf{x})$, flows in the
pseudospin QH ferromagnet. In this paper, we present a microscopic
formulation of the phase current, and show that the phase current arranges
itself to minimize the total energy of the system and makes the Hall
resistance vanish in a counterflow geometry\cite{Kellogg04,Tutuc04}.
Furthermore, it explains also the anomalous Hall resistance in the drag
experiment\cite{Kellogg02}.

This paper is composed as follows. Section II\ is devoted to a concise
review of the microscopic formalism of the QH system based on the
noncommutative geometry. In Section III we analyze the Heisenberg equation
of motion in the QH\ system. In Section IV the formula is derived for the
current in the spin ferromagnet from the continuity equation. In Section V
we study the current in the pseudospin ferromagnet. In Section VI we
determine the phase current by minimizing the total energy of the system. In
Section VII we investigate how the anomalous Hall resistance occurs in the
pseudospin ferromagnet.

\section{Noncommutative Geometry}

A planar electron performs cyclotron motion in magnetic field, $\mathbf{B}%
=(0,0,-B_{\perp })$. The electron coordinate $\mathbf{x}=(x,y)$ is
decomposed into the guiding center $\mathbf{X}=(X,Y)$ and the relative
coordinate $\mathbf{R}=(R_{x},R_{y})$, $\mathbf{x}=\mathbf{X}+\mathbf{R}$,
where $R_{x}=-P_{y}/eB_{\perp }$ and $R_{y}=P_{x}/eB_{\perp }$ with $\mathbf{%
P}=(P_{x},P_{y})$ the covariant momentum. The commutation relations are $%
[X,Y]=-i\ell _{B}^{2}$, $[P_{x},P_{y}]=i{\hbar ^{2}/}\ell _{B}^{2}$ and $%
[X,P_{x}]=[X,P_{y}]=[Y,P_{x}]=[Y,P_{y}]=0$, with $\ell _{B}^{2}=\hbar
/eB_{\perp }$. They imply that the guiding center and the relative
coordinate are independent variables.

The kinetic Hamiltonian,%
\begin{equation}
H_{\text{K}}=\frac{\mathbf{P}^{2}}{2M}=\frac{1}{2M}%
(P_{x}-iP_{y})(P_{x}+iP_{y})+\frac{1}{2}\hbar \omega _{\text{c}},
\label{KinemHamilA}
\end{equation}%
creates Landau levels with gap energy $\hbar \omega _{\text{c}}=\hbar
eB_{\perp }/M$. When it is large enough, excitations across Landau levels
are suppressed at sufficiently low temperature. It is a good approximation
to prohibit all such excitations by requiring electron confinement to the
Lowest Landau level.

We explore the physics of electrons confined to the lowest Landau level,
where the electron position is specified solely by the guiding center $%
\mathbf{X}=(X,Y)$, whose $X$ and $Y$ components are noncommutative,%
\begin{equation}
\lbrack X,Y]=-i\ell _{B}^{2}.  \label{AlgebGC}
\end{equation}%
We introduce the operators 
\begin{equation}
b=\frac{1}{\sqrt{2}\ell _{B}}(X-iY),\quad b^{\dag }=\frac{1}{\sqrt{2}\ell
_{B}}(X+iY),
\end{equation}%
obeying $[b,b^{\dag }]=1$, and define the Fock states 
\begin{equation}
|n\rangle =\frac{1}{\sqrt{n!}}(b^{\dag })^{n}|0\rangle ,\qquad b|0\rangle =0,
\end{equation}%
where $n=0,1,2,\cdots $. The QH system provides us with an ideal
2-dimensional world with the built-in noncommutative geometry.

We assume that the electron carries the SU(N) index. For instance, it
carries the spin SU(2) index in the monolayer system, and the
spin-pseudospin SU(4) index in the bilayer system. The SU(N) electron field $%
\Psi (\mathbf{x})$ has $N$ components. It is given by%
\begin{equation}
\psi _{\mu }(\mathbf{x})=\sum_{n}\langle \mathbf{x}|n\rangle c_{\mu }(n)
\label{ElectField}
\end{equation}%
for electrons in the lowest Landau level, where $c_{\mu }(n)$ is the
annihilation operator acting on the Fock state $|n\rangle $, 
\begin{equation}
\{c_{\mu }(n),c_{\nu }^{\dag }(m)\}=\delta _{mn}\delta _{\mu \nu }.
\label{AlgebFermi}
\end{equation}%
The physical variables are the electron density $\rho (\mathbf{x})$ and the
isospin field $I_{A}(\mathbf{x})$,%
\begin{equation}
\rho (\mathbf{x})=\Psi ^{\dagger }(\mathbf{x})\Psi (\mathbf{x}),\quad I_{A}(%
\mathbf{x})={\frac{1}{2}}\Psi ^{\dagger }(\mathbf{x})\lambda _{A}\Psi (%
\mathbf{x}),
\end{equation}%
where $\lambda _{A}$ are the generating matrices of SU(N). We summarize the
electron density and the isospin density into the density matrix as%
\begin{equation}
D_{\mu \nu }=\frac{1}{N}\delta _{\mu \nu }\rho +(\lambda _{A})_{\mu \nu
}I_{A}.
\end{equation}%
It is given by%
\begin{equation}
D_{\mu \nu }(\mathbf{p})=e^{-\ell _{B}^{2}\mathbf{p}^{2}/4}\hat{D}_{\mu \nu
}(\mathbf{p})
\end{equation}%
in the momentum space, together with%
\begin{equation}
\hat{D}_{\mu \nu }(\mathbf{p})=\frac{1}{2\pi }\sum_{mn}\langle m|e^{-i%
\mathbf{pX}}|n\rangle c_{\nu }^{\dagger }(m)c_{\mu }(n).
\end{equation}%
We call $\hat{D}_{\mu \nu }(\mathbf{p})$ the bare density. The difference
between $D_{\mu \nu }$ and $\hat{D}_{\mu \nu }$ is negligible for
sufficiently smooth field configurations.

In the succeeding analysis of the dynamics of the SU(N) QH system, the W$%
_{\infty }$(N) algebra satisfied by the bare density, 
\begin{align}
2\pi \lbrack \hat{D}_{\mu \nu }(\mathbf{p}),\hat{D}_{\sigma \tau }(\mathbf{q}%
)]=& \delta _{\mu \tau }e^{+\frac{i}{2}\ell _{B}^{2}\mathbf{p}\wedge \mathbf{%
q}}\hat{D}_{\sigma \nu }(\mathbf{p}+\mathbf{q})  \notag \\
& -\delta _{\sigma \nu }e^{-\frac{i}{2}\ell _{B}^{2}\mathbf{p}\wedge \mathbf{%
q}}\hat{D}_{\mu \tau }(\mathbf{p}+\mathbf{q}),  \label{AlgebWpre}
\end{align}%
plays the basic role. We have already derived this relation based on the
noncommutative relation (\ref{AlgebGC}) and the anticommutation relation (%
\ref{AlgebFermi}) in our previous works\cite{EzawaX03B,Ezawa05D}: See (3.19)
in the first reference of Ref.\cite{Ezawa05D}.

\section{Equations of Motion}

The Heisenberg equation of motion determines the quantum mechanical system.
It is given by%
\begin{equation}
i\hbar {\frac{d}{dt}}\hat{D}_{\mu \nu }(\mathbf{p})=[\hat{D}_{\mu \nu }(%
\mathbf{p}),H]  \label{HeiseLLL}
\end{equation}%
for electrons in the lowest Landau level. In ordinary physics the dynamics
arises from the kinetic Hamiltonian. However, this is not the case here,
since the kinetic Hamiltonian (\ref{KinemHamilA}) commutes with $\hat{D}%
_{\mu \nu }(\mathbf{p})$. Indeed, $H_{\text{K}}$ contains only the relative
coordinate $\mathbf{R}=(-P_{y}/eB_{\perp },P_{x}/eB_{\perp })$, while $\hat{D%
}_{\mu \nu }(\mathbf{p})$ contains only the guiding center $\mathbf{X}=(X,Y)$%
. Nontrivial dynamics can arise because the guiding center has the
noncommutative coordinates obeying (\ref{AlgebGC}). It is remarkable that
the dynamics arises from the very nature of noncommutative geometry in the
QH system.

The total Hamiltonian $H$ consists of the Coulomb term $H_{\text{C}}$ and
the rest term $H_{\text{rest}}$, 
\begin{equation}
H=H_{\text{C}}+H_{\text{rest}},  \label{HamilA}
\end{equation}%
where $H_{\text{rest}}$ stands for the Zeeman term in the monolayer system
and additionally the tunneling and bias terms in the bilayer system. All of
them are represented in terms of the bare densities. Hence, we are able to
calculate the Heisenberg equation of motion (\ref{HeiseLLL}) based on the W$%
_{\infty }$(N) algebra (\ref{AlgebWpre}).

What is observed experimentally is the classical field $\mathcal{D}_{\mu \nu
}(\mathbf{x})$, which is the expectation value of $\hat{D}_{\mu \nu }(%
\mathbf{x})$ by a Fock state, 
\begin{equation}
\mathcal{D}_{\mu \nu }(\mathbf{x})=\langle \hat{D}_{\mu \nu }(\mathbf{x}%
)\rangle .
\end{equation}%
We consider the class of Fock states which can be written as 
\begin{equation}
|\mathfrak{S}\rangle =e^{iW}|\mathfrak{S}_{0}\rangle ,  \label{SkyrmFormuC}
\end{equation}%
where $W$ is an arbitrary element of the W$_{\infty }(N)$ algebra which
represents a general linear combination of the operators $c_{\nu }^{\dagger
}(m)c_{\mu }(n)$. The state $|\mathfrak{S}_{0}\rangle $ is assumed to be of
the form\cite{Ezawa05D}%
\begin{equation}
|\mathfrak{S}_{0}\rangle =\prod_{\mu ,n}\left[ c_{\mu }^{\dag }(n)\right]
^{\nu _{\mu }(n)}|0\rangle ,  \label{SkyrmFormuCx}
\end{equation}%
where $\nu _{\mu }(n)$ takes the value either $0$ or $1$ specifying whether
the isospin state $\mu $ at the state $|n\rangle $ is occupied or not,
respectively. Though it may not include all states, it certainly contains
all integer QH states, by which we mean the ground state as well as all
quasiparticle excited states at $\nu =$integer.

The classical field satisfied the classical equation of motion. It is
constructed by taking the expectation value of the Heisenberg equation of
motion (\ref{HeiseLLL}),%
\begin{equation}
i\hbar \frac{d}{dt}\mathcal{D}_{\mu \nu }(\mathbf{x})=\langle \hat{D}_{\mu
\nu }(\mathbf{p})H\rangle -\langle H\hat{D}_{\mu \nu }(\mathbf{p})\rangle .
\end{equation}%
We can verify\cite{Ezawa05D} for the class of states (\ref{SkyrmFormuC}) that%
\begin{equation}
\frac{d}{dt}\mathcal{D}_{\mu \nu }(\mathbf{x})=[\mathcal{D}_{\mu \nu }(%
\mathbf{x}),\mathcal{H}]_{\text{PB}},  \label{EqOfMotionClass}
\end{equation}%
where $\mathcal{H}=\langle H\rangle $ is the classical Hamiltonian, provided
the classical density is endowed with the Poisson structure%
\begin{align}
2\pi i\hbar \lbrack \mathcal{D}_{\mu \nu }(\mathbf{p}),\mathcal{D}_{\sigma
\tau }(\mathbf{q})]_{\text{PB}}=& \delta _{\mu \tau }e^{+\frac{i}{2}\ell
_{B}^{2}\mathbf{p}\wedge \mathbf{q}}\mathcal{D}_{\sigma \nu }(\mathbf{p}+%
\mathbf{q})  \notag \\
& -\delta _{\sigma \nu }e^{-\frac{i}{2}\ell _{B}^{2}\mathbf{p}\wedge \mathbf{%
q}}\mathcal{D}_{\mu \tau }(\mathbf{p}+\mathbf{q}).
\end{align}%
The classical Coulomb energy consists of the direct and exchange energies%
\cite{Ezawa05D}, $\mathcal{H}_{\text{C}}=\mathcal{H}_{\text{D}}+\mathcal{H}_{%
\text{X}}$, and we obtain 
\begin{equation}
\mathcal{H}=\mathcal{H}_{\text{D}}+\mathcal{H}_{\text{X}}+\mathcal{H}_{\text{%
rest}},  \label{HamilB}
\end{equation}%
from the total Hamiltonian (\ref{HamilA}). Here, $\mathcal{H}_{\text{D}}$
and $\mathcal{H}_{\text{rest}}$ have the same expressions as $H_{\text{C}}$
and $H_{\text{rest}}$, respectively, with the replacement of $\hat{D}_{\mu
\nu }$ by $\mathcal{D}_{\mu \nu }$: The exchange term $\mathcal{H}_{\text{X}%
} $ is a new term. We give explicit expressions of these terms in the
following sections.

\section{Currents in Spin QH Ferromagnet}

In the monolayer QH system the physical variables are the electron density $%
\rho (\mathbf{x})$ and the spin field $S_{a}(\mathbf{x})$,%
\begin{equation}
\rho (\mathbf{x})=\Psi ^{\dagger }(\mathbf{x})\Psi (\mathbf{x}),\quad S_{a}(%
\mathbf{x})={\frac{1}{2}}\Psi ^{\dagger }(\mathbf{x})\tau _{a}\Psi (\mathbf{x%
})
\end{equation}%
with (\ref{ElectField}) for $\Psi (\mathbf{x})$. We denote the corresponding
classical field as%
\begin{equation}
\varrho (\mathbf{x})=\langle \hat{\rho}(\mathbf{x})\rangle ,\quad \mathcal{S}%
(\mathbf{x})=\langle \hat{S}(\mathbf{x})\rangle .  \label{ClassS}
\end{equation}%
Here $\tau _{a}$ are the Pauli matrices for the spin space.

We study the electric current in the QH state. The current is introduced
originally to guarantee the charge conservation. This is the case also in
the noncommutative plane,%
\begin{equation}
-e\frac{d}{dt}\hat{\rho}(\mathbf{x})=\partial _{i}\hat{J}_{i}(\mathbf{x}),
\label{CurreA}
\end{equation}%
where $-e\hat{\rho}(\mathbf{x})$ is the charge density and $\hat{J}_{i}(%
\mathbf{x})$ is the current in the lowest Landau level. The physically
observed current is the classical current given by%
\begin{equation}
\mathcal{J}_{i}(\mathbf{x})=\langle \hat{J}_{i}(\mathbf{x})\rangle .
\end{equation}%
Taking the expectation value of (\ref{CurreA}) and using the classical
equation of motion (\ref{EqOfMotionClass}), we have%
\begin{equation}
\partial _{i}\mathcal{J}_{i}(\mathbf{x})=-e\frac{d}{dt}\varrho (\mathbf{x}%
)=-e[\varrho (\mathbf{x}),\mathcal{H}]_{\text{PB}}.  \label{HallCurreMonoB}
\end{equation}%
The formula for the current $\mathcal{J}_{i}(\mathbf{x})$ is obtained from
this continuity equation by integrating it.

In the monolayer QH system the Hamiltonian consists of the Coulomb term, the
Zeeman term and the electric-field term,%
\begin{equation}
H=H_{\text{C}}+H_{\text{Z}}+H_{\text{E}}.
\end{equation}%
We introduce the scalar potential $\varphi (\mathbf{x})$ to produce an
electric field, 
\begin{equation}
E_{i}(\mathbf{x})=-\partial _{i}\varphi (\mathbf{x}),
\end{equation}%
in the Hamiltonian $H_{\text{E}}$. The classical Hamiltonian is given by $%
\mathcal{H}=\langle H\rangle $, which reads\cite{Ezawa05D} 
\begin{equation}
\mathcal{H}=\mathcal{H}_{\text{D}}+\mathcal{H}_{\text{X}}+\mathcal{H}_{\text{%
Z}}+\mathcal{H}_{\text{E}},  \label{ClassHamilMono}
\end{equation}%
where\beginABC\label{ClassHamilSpin}%
\begin{align}
\mathcal{H}_{\text{D}}=& {\pi }\int \!d^{2}q\,V_{\text{D}}(\mathbf{q}%
)\varrho (-\mathbf{q})\varrho (\mathbf{q}),  \label{HamilClassD} \\
\mathcal{H}_{\text{X}}=& -\pi \int d^{2}p\,V_{\text{X}}(\mathbf{p})\Big[%
\sum_{a=xyz}\mathcal{S}_{a}(-\mathbf{p})\mathcal{S}_{a}(\mathbf{p})  \notag
\\
& \hspace{2.5cm}+\frac{1}{4}\varrho (-\mathbf{p})\varrho (\mathbf{p})\Big],
\label{HamilClassX} \\
\mathcal{H}_{\text{Z}}=& -2\pi \Delta _{\text{Z}}\mathcal{S}_{z}(0).
\label{HamilClassZ} \\
\mathcal{H}_{\text{E}}=& -e\int d^{2}q\,e^{-\mathbf{q}^{2}\ell
_{B}^{2}/4}\varphi (-\mathbf{q})\varrho (\mathbf{q}),  \label{HamilClassE}
\end{align}%
\endABC with 
\begin{align}
V_{\text{D}}(\mathbf{q})=& \frac{e^{2}}{4\pi \varepsilon |\mathbf{q}|}%
e^{-\ell _{B}^{2}\mathbf{q}^{2}/2}, \\
V_{\text{X}}(\mathbf{p})=& \frac{\sqrt{2\pi }e^{2}\ell _{B}}{4\pi
\varepsilon }I_{0}(\ell _{B}^{2}\mathbf{p}^{2}/4)e^{-\ell _{B}^{2}\mathbf{p}%
^{2}/4}.
\end{align}%
Here, $I_{0}(x)$ is the modified Bessel function.

It is straightforward to calculate the Poisson bracket (\ref{HallCurreMonoB}%
) with (\ref{ClassHamilMono}),\beginABC%
\begin{align}
& [\varrho (\mathbf{k}),\mathcal{H}_{\text{D}}]_{\text{PB}}  \notag \\
=& \frac{2}{\hbar }\int \!d^{2}q\,V_{\text{D}}(\mathbf{q})\{\varrho (\mathbf{%
q}),{\varrho }(\mathbf{k}-\mathbf{q})\}\sin \left( \ell _{B}^{2}{\frac{%
\mathbf{k}\!\wedge \!\mathbf{q}}{2}}\right) ,  \label{PoissoA} \\
& [\varrho (\mathbf{k}),\mathcal{H}_{\text{X}}]_{\text{PB}}  \notag \\
=& -\frac{1}{2\hbar }\int \!d^{2}q\,V_{\text{X}}(\boldsymbol{q})\varrho (-%
\boldsymbol{q})\varrho (\boldsymbol{k}+\boldsymbol{q})\sin \left( \ell
_{B}^{2}\frac{\mathbf{k}\!\wedge \!\mathbf{q}}{2}\right)  \notag \\
& -\frac{2}{\hbar }\int \!d^{2}q\,V_{\text{X}}(\boldsymbol{q})\emph{S}_{a}(-%
\boldsymbol{q})\emph{S}_{a}(\boldsymbol{k}+\boldsymbol{q})\sin \left( \ell
_{B}^{2}\frac{\mathbf{k}\!\wedge \!\mathbf{q}}{2}\right)  \label{PoissoB} \\
& [\varrho (\mathbf{k}),\mathcal{H}_{\text{Z}}]_{\text{PB}}=0, \\
& [\varrho (\mathbf{k}),\mathcal{H}_{\text{E}}]_{\text{PB}}  \notag \\
=& -\frac{e}{2\pi \hbar }\int \!d^{2}q\,e^{-\mathbf{q}^{2}\ell
_{B}^{2}/4}\varphi (\mathbf{q})\varrho (\mathbf{k}-\mathbf{q})\sin \left(
\ell _{B}^{2}{\frac{\mathbf{k}\!\wedge \!\mathbf{q}}{2}}\right) .
\label{PoissoC}
\end{align}%
\endABC We evaluate them on the ground state in the QH system, where the
classical density is given by 
\begin{equation}
\varrho (\mathbf{k})=2\pi \rho _{0}\delta ^{2}(\mathbf{k})
\label{CondiIncom}
\end{equation}%
with $\rho _{0}$ the total electron density. This is known as the
incompressibility condition\cite{EzawaBook}, implying that the QH system is
an incompressible liquid.

We first examine the Poisson bracket (\ref{PoissoB}) for $\mathcal{H}_{\text{%
D}}$. Substituting the incompressibility condition into (\ref{PoissoA}) it
is trivial to see that $[\varrho (\mathbf{k}),\mathcal{H}_{\text{D}}]_{\text{%
PB}}=0$. Hence, there is no contribution to the Hall current from the direct
Coulomb term $\mathcal{H}_{\text{D}}$.

We next examine the Poisson bracket (\ref{PoissoB}) for $\mathcal{H}_{\text{X%
}}$. The term involving $\varrho (-\boldsymbol{k}^{\prime })\varrho (%
\boldsymbol{k}+\boldsymbol{k}^{\prime })\sin [\ell _{B}^{2}(\mathbf{k}%
\!\wedge \!\mathbf{q})/2]$ vanishes because of the incompressibility
condition. The remaining term involves $\emph{S}_{z}(-\boldsymbol{k}^{\prime
})\emph{S}_{z}(\boldsymbol{k}+\boldsymbol{k}^{\prime })$. Since we are
concerned about a homogeneous flow of electrons, taking the nontrivial
lowest order term in the derivative expansion of potential $V_{\text{X}}(%
\boldsymbol{k})$, we find 
\begin{align}
& \int \!d^{2}q\,V_{\text{X}}(\boldsymbol{k}^{\prime })\emph{S}_{z}(-\mathbf{%
q})\emph{S}_{z}(\boldsymbol{k}+\mathbf{q})\sin \left( \frac{1}{2}\ell
_{B}^{2}\boldsymbol{k}\wedge \mathbf{q}\right)  \notag \\
\simeq & V_{\text{X}}(0)\int \!d^{2}q\,\emph{S}_{z}(-\mathbf{q})\emph{S}_{z}(%
\boldsymbol{k}+\mathbf{q})\sin \left( \frac{1}{2}\ell _{B}^{2}\boldsymbol{k}%
\wedge \mathbf{q}\right)  \notag \\
=& 0.
\end{align}%
This is zero because the relation%
\begin{align}
& \int \!d^{2}q\,f(-\mathbf{q})g(\boldsymbol{k}+\mathbf{q})\sin \left( 
\boldsymbol{k}\wedge \mathbf{q}\right)  \notag \\
& \qquad =-\int \!d^{2}q\,f(\boldsymbol{k}+\mathbf{q})g(-\mathbf{q})\sin
\left( \boldsymbol{k}\wedge \mathbf{q}\right)
\end{align}%
holds for any two functions $f$ and $g$. Hence, there is no contribution
from $\mathcal{H}_{\text{X}}$.

We finally examine the contribution from $\mathcal{H}_{\text{E}}$ by
evaluating (\ref{PoissoC}). Expanding $\sin [\ell _{B}^{2}(\mathbf{k}%
\!\wedge \!\mathbf{q})/2{]}$, we obtain 
\begin{align}
\mathcal{J}_{i}(\mathbf{k})=& -i\frac{e^{2}\ell _{B}^{2}}{2\pi \hbar }%
\varepsilon _{ij}\int \!d^{2}q\,e^{-\mathbf{k}^{2}\ell
_{B}^{2}/4}q_{j}\varphi (\mathbf{q})\varrho (\mathbf{k}-\mathbf{q})  \notag
\\
& \qquad \times \Big[1-\frac{1}{3!}\Big(\ell _{B}^{2}{\frac{\mathbf{k}%
\!\wedge \!\mathbf{q}}{2}}\Big)^{2}+\cdots \Big].
\end{align}%
In a constant electric field $E_{j}$ such that%
\begin{equation}
q_{i}\varphi (\mathbf{q})=2\pi iE_{j}\delta (\mathbf{q}),
\end{equation}%
we find%
\begin{equation}
\mathcal{J}_{i}(\mathbf{k})=\frac{e^{2}}{\hbar }\ell _{B}^{2}\varepsilon
_{ij}E_{j}e^{-\mathbf{k}^{2}\ell _{B}^{2}/4}\varrho (\mathbf{k}-\mathbf{q}).
\end{equation}%
On the incompressible state (\ref{CondiIncom}) it yields 
\begin{equation}
\mathcal{J}_{i}(\mathbf{x})=\frac{e^{2}\ell _{B}^{2}}{\hbar }\varepsilon
_{ij}E_{j}\rho _{0}.  \label{HallCurreMonoD}
\end{equation}%
This is the standard formula for the Hall\ current.

We have demonstrated that the Coulomb and Zeeman interactions do not affect
the Hall current, as expected. However, this is not a trivial result since
the exchange Coulomb interaction yields rather complicated formulas in the
midstream of calculations. As we have verified, they vanish in the spin QH
ferromagnet. On the other hand, as we shall see in the succeeding sections,
the exchange Coulomb interaction produces the phase current in the
pseudospin QH ferromagnet.

\section{Currents in Pseudospin QH Ferromagnet}

\label{SecElectCurreBL}

We proceed to study the electric currents in the bilayer system. Though the
actual system has the spin-pseudospin SU(4) structure, we consider the
spin-frozen system since the spin does not affect the current: See Appendix
B. The electron field (\ref{ElectField}) has two components $\psi ^{\alpha }(%
\mathbf{x})$ corresponding to the front ($\alpha $=f) and back ($\alpha $=b)
layers. The physical variables are the electron densities $\rho ^{\alpha }(%
\mathbf{x})$ in the two layers, and the pseudospin field $P_{a}(\mathbf{x})$,%
\begin{equation}
\rho ^{\alpha }(\mathbf{x})=\psi ^{\alpha \dagger }(\mathbf{x})\psi ^{\alpha
}(\mathbf{x}),\quad P_{a}(\mathbf{x})={\frac{1}{2}}\Psi ^{\dagger }(\mathbf{x%
})\pi _{a}\Psi (\mathbf{x})
\end{equation}%
with (\ref{ElectField}), where $\pi _{a}$ are the Pauli matrices for the
pseudospin space. We use notations%
\begin{equation}
\varrho ^{\alpha }(\mathbf{x})=\langle \hat{\rho}^{\alpha }(\mathbf{x}%
)\rangle ,\quad \mathcal{P}_{a}(\mathbf{x})=\langle \hat{P}_{a}(\mathbf{x}%
)\rangle  \label{ClassP}
\end{equation}%
for the classical variables.

The Hamiltonian $H$ consists of the Coulomb term, the tunneling term, the
gate term and the electric-field term,%
\begin{equation}
H=H_{\text{C}}+H_{\text{T}}+H_{\text{gate}}+H_{\text{E}},
\end{equation}%
which are explicitly given by (\ref{BLHamils}) in Appendix A. The gate term
has been introduced to make a density imbalance between the two layers. The
average density reads%
\begin{equation}
\rho _{0}^{\text{f}}=\frac{1+\sigma _{0}}{2}\rho _{0},\text{\quad }\rho
_{0}^{\text{b}}=\frac{1-\sigma _{0}}{2}\rho _{0}  \label{DensiFB}
\end{equation}%
in each layer. We call $\sigma _{0}$ the imbalance parameter.

As we show in Appendix A, the classical Hamiltonian $\mathcal{H}$ is
rearranged into 
\begin{equation}
\mathcal{H}\equiv \langle H\rangle =\mathcal{H}_{\text{D}}+\mathcal{H}_{%
\text{X}}+\mathcal{H}_{\text{T}}+\mathcal{H}_{\text{bias}}+\mathcal{H}_{%
\text{E}},  \label{HamilBL}
\end{equation}%
where $\mathcal{H}_{\text{D}}$ and $\mathcal{H}_{\text{X}}$ are the direct
and exchange Coulomb energy terms,\beginABC\label{ClassHamilBL} 
\begin{align}
\mathcal{H}_{\text{D}}& =\pi \int d^{2}pV_{\text{D}}^{+}(\boldsymbol{p}%
)\varrho (-\boldsymbol{p})\varrho (\boldsymbol{p})  \notag \\
& +4\pi \int d^{2}pV_{\text{D}}^{-}(\boldsymbol{p})\mathcal{P}_{z}(-%
\boldsymbol{p})\mathcal{P}_{z}(\boldsymbol{p})-8\pi \epsilon _{\text{D}%
}^{-}\sigma _{0}\mathcal{P}_{z}(0) \\
\mathcal{H}_{\text{X}}& =-\pi \sum_{a=x,y}\int d^{2}pV_{\text{X}}^{d}(%
\boldsymbol{p})\mathcal{P}_{a}(-\boldsymbol{p})\mathcal{P}_{a}(\boldsymbol{p}%
)  \notag \\
& -\pi \int d^{2}pV_{\text{X}}(\boldsymbol{p})\bigg[\mathcal{P}_{z}(-%
\boldsymbol{p})\mathcal{P}_{z}(\boldsymbol{p})+\frac{1}{4}\varrho (-%
\boldsymbol{p})\varrho (\boldsymbol{p})\bigg]  \notag \\
& +8\pi \epsilon _{\text{X}}^{-}\sigma _{0}\mathcal{P}_{z}(0),
\end{align}%
and $\mathcal{H}_{\text{T}}$ and $\mathcal{H}_{\text{bias}}$ are the
tunneling and bias terms, 
\begin{align}
\mathcal{H}_{\text{T}}& =-2\pi \Delta _{\text{SAS}}{\emph{P}}_{x}(0),
\label{HamilClassT} \\
\mathcal{H}_{\text{bias}}& =-2\pi \frac{\sigma _{0}}{\sqrt{1-\sigma _{0}^{2}}%
}\Delta _{\text{SAS}}\emph{P}_{z}(0),  \label{HamilClassB}
\end{align}%
while the electric-field term reads%
\begin{equation}
\mathcal{H}_{\text{E}}=-e\int \!d^{2}q\,e^{-\mathbf{q}^{2}\ell _{B}^{2}/4}%
\left[ \varphi ^{\text{f}}(-\mathbf{q})\varrho ^{\text{f}}(\mathbf{q}%
)+\varphi ^{\text{b}}(-\mathbf{q})\varrho ^{\text{b}}(\mathbf{q})\right] .
\label{EfieldBL}
\end{equation}%
\endABC Various Coulomb potentials are defined by%
\begin{equation}
V_{\text{X}}=V_{\text{X}}^{+}+V_{\text{X}}^{-},\qquad V_{\text{X}}^{d}=V_{%
\text{X}}^{+}-V_{\text{X}}^{-},
\end{equation}%
and\beginABC%
\begin{align}
V_{\text{D}}^{\pm }(\mathbf{p})=& {\frac{e^{2}}{8\pi \varepsilon |\mathbf{p}|%
}}\left( {1\pm }e^{-|\mathbf{p}|d}\right) e^{-\frac{1}{2}\ell _{B}^{2}%
\mathbf{p}^{2}}, \\
V_{\text{X}}^{\pm }(\mathbf{p})=& \frac{\ell _{B}^{2}}{\pi }\int
\!d^{2}k\,e^{-i\ell _{B}^{2}\mathbf{p}\wedge \mathbf{k}}V_{\text{D}}^{\pm }(%
\mathbf{k}),
\end{align}%
\endABC with the interlayer separation $d$. We have also defined%
\begin{equation}
\epsilon _{\text{D}}^{-}=\frac{1}{2}\rho _{0}\int {\!}d^{2}x{\,}V_{\text{D}%
}^{-}(\mathbf{x}),\quad \epsilon _{\text{X}}^{-}=\frac{1}{4}\rho _{0}\int {\!%
}d^{2}x{\,}V_{\text{X}}^{-}(\mathbf{x}).  \label{ParamE}
\end{equation}%
We note that the capacitance energy is given by 
\begin{equation}
\epsilon _{\text{cap}}=4(\epsilon _{\text{D}}^{-}-\epsilon _{\text{X}}^{-}).
\label{CapacParam}
\end{equation}%
The scalar potentials $\varphi ^{\text{f}}(\mathbf{x})$ and $\varphi ^{\text{%
b}}(\mathbf{x})$ are introduced to produce the electric fields%
\begin{equation}
E_{i}^{\text{f}}=-\partial _{i}\varphi ^{\text{f}},\qquad E_{i}^{\text{b}%
}=-\partial _{i}\varphi ^{\text{b}}
\end{equation}%
in the front and back layers within the Hamiltonian (\ref{EfieldBL}).

We investigate the classical equation of motion (\ref{EqOfMotionClass}) to
derive the classical current $\mathcal{J}_{i}^{\alpha }(\mathbf{x})$ in each
layer. It is defined so that the charge conserves locally,\beginABC%
\begin{align}
-e\frac{d\varrho ^{\text{f}}}{dt}=& \partial _{i}\mathcal{J}_{i}^{\text{f}}(%
\mathbf{x})-\frac{1}{d}\mathcal{J}_{z}(\mathbf{x}), \\
-e\frac{d\varrho ^{\text{b}}}{dt}=& \partial _{i}\mathcal{J}_{i}^{\text{b}}(%
\mathbf{x})+\frac{1}{d}\mathcal{J}_{z}(\mathbf{x}),
\end{align}%
\endABC where $\mathcal{J}_{z}(\mathbf{x})$ is the tunneling current between
the two layers.

The tunneling term $\mathcal{H}_{\text{T}}$ contributes only to the
tunneling current $\mathcal{J}_{z}(\mathbf{x})$, and it is given by%
\begin{equation}
\frac{1}{d}\mathcal{J}_{z}(\mathbf{x})=\left. \frac{e}{2}\frac{d}{dt}[{%
\varrho }^{\text{f}}-\varrho ^{\text{b}}]\right\vert _{\mathcal{H}_{\text{T}%
}}=e[\mathcal{P}_{z},\mathcal{H}_{\text{T}}]_{\text{PB}}.
\end{equation}%
The current in the layer $\alpha =$f, b is given by%
\begin{equation}
\partial _{i}\mathcal{J}_{i}^{\alpha }(\mathbf{x})=-e\left. \frac{d\varrho
^{\alpha }}{dt}\right\vert _{\mathcal{H}_{\text{D}}+\mathcal{H}_{\text{X}}+%
\mathcal{H}_{\text{bias}}+\mathcal{H}_{\text{E}}},
\end{equation}%
which consists of $[\varrho ^{\alpha },\mathcal{H}_{\text{D}}]_{\text{PB}}$, 
$[\varrho ^{\alpha },\mathcal{H}_{\text{X}}]_{\text{PB}}$, $[\varrho
^{\alpha },\mathcal{H}_{\text{bias}}]_{\text{PB}}$ and $[\varrho ^{\alpha },%
\mathcal{H}_{\text{E}}]_{\text{PB}}$. We express $\varrho ^{\alpha }$ in
terms of $\varrho $ and $\mathcal{P}_{z}$ as%
\begin{equation}
\varrho ^{\text{f}}=\frac{1}{2}\varrho +\mathcal{P}_{z},\qquad \varrho ^{%
\text{b}}=\frac{1}{2}\varrho -\mathcal{P}_{z}.
\end{equation}%
Note that $\varrho ^{\alpha }=\rho _{0}^{\alpha }$ with (\ref{DensiFB}) in
the ground state.

We need to calculate the Poisson brackets for $\varrho $ and $\mathcal{P}%
_{z} $ with various Hamiltonians. When we estimate them on the state
satisfying the incompressibility condition (\ref{CondiIncom}), many terms
vanish precisely by the same reasons as in the monolayer system. We now
demonstrate that there exists a new contribution from the exchange
interaction $\mathcal{H}_{\text{X}}$ to the current, as is the novel feature
in the pseudospin QH ferromagnet. Let us explicitly write down only those
parts in the Poisson brackets that yield nonvanishing contributions to the
current.

There is a new term involving $\cos (\frac{1}{2}\ell _{B}^{2}\boldsymbol{k}%
\wedge \mathbf{q)}$ in the Poisson bracket $[\mathcal{P}_{z}(\boldsymbol{k}),%
\mathcal{H}_{\text{X}}]_{\text{PB}}$,

\begin{align}
\lbrack & \mathcal{P}_{z}(\boldsymbol{k}),\mathcal{H}_{\text{X}}]_{\text{PB}}
\notag \\
=& -\frac{\epsilon _{ab}}{\hbar }\int \!d^{2}q\,V_{\text{X}}^{d}(\boldsymbol{%
q})\mathcal{P}_{a}(-\boldsymbol{q})\mathcal{P}_{b}(\boldsymbol{k}+%
\boldsymbol{q})\cos \left( \frac{\ell _{B}^{2}}{2}\boldsymbol{k\!}\wedge \!%
\boldsymbol{q}\right)  \notag \\
& +\cdots ,  \label{PBPzX}
\end{align}%
where the index $a$ runs over $x$ and $y$. Making the derivative expansion
of $V_{\text{X}}^{d}(\boldsymbol{q})$ and $\cos (\frac{1}{2}\ell _{B}^{2}%
\boldsymbol{k}\wedge \mathbf{q)}$, we find%
\begin{equation}
\mathcal{J}_{i}^{\text{f(X)}}(\mathbf{x})=-\mathcal{J}_{i}^{\text{b(X)}}(%
\mathbf{x})=\frac{4eJ_{s}^{d}}{\hbar \rho _{0}^{2}}(\partial _{i}\mathcal{P}%
_{x}\cdot \mathcal{P}_{y}-\partial _{i}\mathcal{P}_{y}\cdot \mathcal{P}_{x}),
\label{BLHallX}
\end{equation}%
where%
\begin{equation}
J_{s}^{d}=J_{s}\Big[-\sqrt{\frac{2}{\pi }}\frac{d}{\ell _{B}}+\Big(1+{\frac{%
d^{2}}{\ell _{B}^{2}}}\Big)e^{d^{2}/2\ell _{B}^{2}}\text{erfc}\left( d/\sqrt{%
2}\ell _{B}\right) \Big]  \label{StiffPM}
\end{equation}%
together with%
\begin{equation}
J_{s}={\frac{1}{16\sqrt{2\pi }}}\frac{e^{2}}{4\pi \varepsilon \ell _{B}}
\label{StiffParam}
\end{equation}%
is the pseudospin stiffness.

The electric field yields a nonzero contribution as in the monolayer case,%
\begin{align}
& [\varrho ^{\alpha }(\boldsymbol{k}),\mathcal{H}_{\text{E}}]_{\text{PB}} 
\notag \\
=& -\frac{e}{2\pi \hbar }\int \!d^{2}q\,e_{0}^{-\frac{1}{4}\ell
_{B}^{2}q}\varphi ^{\alpha }(\boldsymbol{q})\varrho ^{\alpha }(\boldsymbol{k}%
-\boldsymbol{q})\sin \left( \ell _{B}^{2}\frac{\boldsymbol{k}\wedge 
\boldsymbol{q}}{2}\right) ,  \notag
\end{align}%
as corresponds to the monolayer formula (\ref{HallCurreMonoD}). Hence, we
obtain the standard formula for the Hall\ current in each layer,%
\begin{equation}
\mathcal{J}_{i}^{\alpha \text{(E)}}(\mathbf{x})=\frac{e^{2}\ell _{B}^{2}}{%
\hbar }\varepsilon _{ij}E_{j}\rho _{0}^{\alpha },  \label{BLHallE}
\end{equation}%
on the incompressible ground state.

Finally, the Poisson bracket with the tunneling Hamiltonian is exactly
calculable, 
\begin{equation}
\lbrack \emph{P}_{y}(\boldsymbol{k}),\mathcal{H}_{\text{T}}]_{\text{PB}}=%
\frac{1}{\hbar }\Delta _{\text{SAS}}\emph{P}_{z}(\boldsymbol{k}),
\end{equation}%
which yields 
\begin{equation}
\mathcal{J}_{z}(\mathbf{x})=-\frac{ed}{\hbar }\Delta _{\text{SAS}}\emph{P}%
_{y}(\mathbf{x})
\end{equation}%
to the tunneling current.

We parametrize the classical pseudospin field in terms of the interlayer
phase field $\vartheta (\mathbf{x})$ and the imbalance field $\sigma (%
\mathbf{x})$,%
\begin{align}
& \emph{P}_{x}(\mathbf{x})={\frac{1}{2}\rho }_{0}\sqrt{1-\sigma ^{2}(\mathbf{%
x})}\cos \vartheta (\mathbf{x}),  \notag \\
& \emph{P}_{y}(\mathbf{x})=-{\frac{1}{2}\rho }_{0}\sqrt{1-\sigma ^{2}(%
\mathbf{x})}\sin \vartheta (\mathbf{x}),  \notag \\
& \emph{P}_{z}(\mathbf{x})={\frac{1}{2}\rho }_{0}\sigma (\mathbf{x}).
\label{PpinParam}
\end{align}%
The Hall current is given by the sum of (\ref{BLHallX}) and (\ref{BLHallE}),%
\beginABC\label{CurreBLa}%
\begin{align}
\mathcal{J}_{i}^{\text{f}}(\mathbf{x})=& \frac{eJ_{s}^{d}}{\hbar }(1-\sigma
^{2})\partial _{i}\vartheta +\frac{e^{2}\ell _{B}^{2}}{\hbar }\varepsilon
_{ij}E_{j}^{\text{f}}\varrho ^{\text{f}}, \\
\mathcal{J}_{i}^{\text{b}}(\mathbf{x})=& -\frac{eJ_{s}^{d}}{\hbar }(1-\sigma
^{2})\partial _{i}\vartheta +\frac{e^{2}\ell _{B}^{2}}{\hbar }\varepsilon
_{ij}E_{j}^{\text{b}}\varrho ^{\text{b}}.
\end{align}%
\endABC We have shown that the phase current arises in the presence of the
interlayer phase coherence.

We should mention that the emergence of the phase current, $\propto \partial
_{i}\vartheta (\mathbf{x})$, in the pseudospin QH ferromagnet has already
been pointed out in an effective theory\cite{EzawaBook} based on an
intuitive and phenomenological reasoning. In this paper we have presented a
microscopic formulation of the phase current.

\begin{figure}[t]
\includegraphics[width=0.42\textwidth]{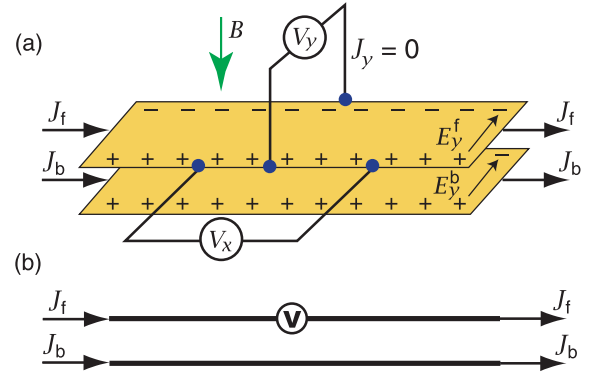}
\caption{ (a) Hall currents are injected to the front and back layers
independently. The Hall and diagonal resistances are measured only in one of
the layers. (b) A simplified picture is given to represent the same
measurement, where the symbol V in a circle indicates that the measurement
is done on this layer.}
\label{FigHallCurBL}
\end{figure}

\section{Diagonal and Hall Resistances}

\label{SecBLQHcurre}

Let us first review the Hall current in the monolayer system with
homogeneous electron density $\rho _{0}$. The electric field $\mathbf{E}$
drives the Hall current into the direction perpendicular to it,%
\begin{equation}
\mathcal{J}_{i}=\frac{e^{2}\ell _{B}^{2}\rho _{0}}{\hbar }\varepsilon
_{ij}E_{j}.  \label{CurreQH}
\end{equation}%
We apply the electric field so that the current flows into the $x$
direction, as implies $E_{x}=0$. Hence the diagonal resistance vanishes,%
\begin{equation}
R_{xx}\equiv {\frac{E_{x}}{\mathcal{J}_{x}}}=0,  \label{DispaLess}
\end{equation}%
and the Hall resistance is given by%
\begin{equation}
R_{xy}\equiv {\frac{E_{y}}{\mathcal{J}_{x}}}=\frac{\hbar }{e^{2}\ell
_{B}^{2}\rho _{0}}=\frac{2\pi \hbar }{\nu e^{2}}.
\end{equation}%
The signals of the QH effect consist of the dissipationless current (\ref%
{DispaLess}) and the development of the Hall plateau at the magic filling
factor $\nu $.

What occurs in actual systems is as follows. We feed the current $\mathcal{J}%
_{x}$ into the $x$ direction. Due to the Lorentz force electrons accumulate
at the edge of the sample, which generates such an electric field $E_{y}$
that makes the given amount of current $\mathcal{J}_{x}$ flow into the $x$
direction [Fig.\ref{FigHallCurBL}]. The relation between the current and the
electric field is fixed kinematically by the formula (\ref{CurreQH}) in the
monolayer system.

We proceed to study the QH current in the imbalanced bilayer system at $%
\sigma _{0}$, where the electron densities are given by (\ref{DensiFB}) in
the ground state. We assume the sample parameter $\Delta _{\text{SAS}}=0$ so
that there is no tunneling current between the two layers. As we show in
Appendix A, the interlayer phase field $\vartheta (\mathbf{x})$ is gapless
in the limit $\Delta _{\text{SAS}}=0$, but the imbalance field $\sigma (%
\mathbf{x})$ has the gap $\epsilon _{\text{cap}}$. Consequently, the
excitation of $\sigma (\mathbf{x})$ is suppressed at sufficiently low
temperature. Hence we set $\sigma (\mathbf{x})=\sigma _{0}$ in all formulas.

We imagine the electric fields $E_{j}^{\text{f}}$ and $E_{j}^{\text{b}}$
driving the Hall currents to flow into the $x$ direction [Fig.\ref%
{FigHallCurBL}]. As we have argued in the previous section, the basic
formula for the current is (\ref{CurreBLa}), or\beginABC%
\begin{align}
\mathcal{J}_{i}^{\text{f}}=& \frac{e}{\hbar }(1-\sigma
_{0}^{2})J_{s}^{d}\partial _{i}\vartheta +\frac{e^{2}\ell _{B}^{2}}{\hbar }%
\varepsilon _{ij}E_{j}^{\text{f}}\rho _{0}^{\text{f}}, \\
\mathcal{J}_{i}^{\text{b}}=& -\frac{e}{\hbar }(1-\sigma
_{0}^{2})J_{s}^{d}\partial _{i}\vartheta +\frac{e^{2}\ell _{B}^{2}}{\hbar }%
\varepsilon _{ij}E_{j}^{\text{b}}\rho _{0}^{\text{b}},
\end{align}%
\endABC in the imbalance configuration at $\sigma _{0}$.

Since our system is assumed to be homogeneous in the $y$ direction, the
variables depend only on $x$. Thus,%
\begin{equation}
E_{x}^{\text{f}}=E_{x}^{\text{b}}=0.\qquad \partial _{y}\vartheta =0,
\end{equation}%
and%
\begin{equation}
R_{xx}^{\text{f}}\equiv {\frac{E_{x}^{\text{f}}}{\mathcal{J}_{x}^{\text{f}}}}%
=0,\qquad R_{xx}^{\text{b}}\equiv {\frac{E_{x}^{\text{b}}}{\mathcal{J}_{x}^{%
\text{b}}}}=0,
\end{equation}%
for $\mathcal{J}_{x}^{\text{f}}\neq 0$ and $\mathcal{J}_{x}^{\text{b}}\neq 0$%
. The Hall current is given by\beginABC\label{HallCurre}%
\begin{align}
\mathcal{J}_{x}^{\text{f}}=& \frac{e}{\hbar }(1-\sigma
_{0}^{2})J_{s}^{d}\partial _{x}\vartheta +\frac{e^{2}\ell _{B}^{2}\rho _{0}}{%
2\hbar }(1+\sigma _{0})E_{y}^{\text{f}}, \\
\mathcal{J}_{x}^{\text{b}}=& -\frac{e}{\hbar }(1-\sigma
_{0}^{2})J_{s}^{d}\partial _{x}\vartheta +\frac{e^{2}\ell _{B}^{2}\rho _{0}}{%
2\hbar }(1-\sigma _{0})E_{y}^{\text{f}}.
\end{align}%
\endABC Consequently the relation between the current and the electric field
is not fixed kinematically in the presence of the interlayer phase
difference $\vartheta (\mathbf{x})$.

Any set of $E_{y}^{\text{f}}$, $E_{y}^{\text{b}}$ and $\vartheta $ seems to
yield the given amounts of currents $\mathcal{J}_{x}^{\text{f}}$ and $%
\mathcal{J}_{x}^{\text{b}}$ provided they satisfy (\ref{HallCurre}). In the
actual system the unique set of them is realized: It is the one that
minimizes the energy of the system. It is a dynamical problem how $E_{y}^{%
\text{f}}$ and $E_{y}^{\text{b}}$ are determined in the bilayer system.

A bilayer system consists of the two layers and the volume between them. The
dynamics of electrons is described by the Hamiltonian (\ref{HamilBL}), which
is defined on the two planes. The tunneling term has been introduced just to
guarantee the charge conservation. We have so far neglected the electric
field in the volume between the two layers, since it does not contribute to
the equation of motion (\ref{EqOfMotionClass}) for electrons. We now need to
analyze the equation of motion also for the electric field. Equivalently it
is necessary to minimize the Coulomb energy stored in the volume between the
two layers.

The energy due to the electric field $\mathbf{E}(\mathbf{x},z)$ between the
two layers is given by the sum of the Maxwell term and the source term, 
\begin{equation}
H_{\text{E}}={\frac{\varepsilon }{2}}\int d^{2}xdz\,\mathbf{E}^{2}(\mathbf{x}%
,z)-e\int d^{2}xdz\,\varphi (\mathbf{x},z)\rho (\mathbf{x},z),
\label{MaxweBLa}
\end{equation}%
where%
\begin{equation}
E_{x}=-\partial _{x}\varphi ,\quad E_{y}=\partial _{y}\varphi ,\quad E_{z}=-%
\frac{\varphi ^{\text{f}}-\varphi ^{\text{b}}}{d}.
\end{equation}%
Note that (\ref{MaxweBLa}) is equivalent to the "surface term" (\ref%
{EfieldBL}) in the equation of motion (\ref{EqOfMotionClass}) since the
electron density $\rho (\mathbf{x},z)$ is nonzero only on the two layers.

When there are constant fields $E_{y}^{\text{f}}$ and $E_{y}^{\text{b}}$
into the $y$ direction on the layers, the field $E_{z}$ between the two
layers is given by%
\begin{equation}
E_{z}=-\frac{\varphi ^{\text{f}}-\varphi ^{\text{b}}}{d}=\frac{E_{y}^{\text{f%
}}-E_{y}^{\text{b}}}{d}y.
\end{equation}%
We carry out the integration over $z$ and then over the plane in (\ref%
{MaxweBLa}),%
\begin{align}
H_{\text{E}}=& {\frac{\varepsilon d_{w}}{2}}\int d^{2}x\,\left( (E_{y}^{%
\text{f}})^{2}+(E_{y}^{\text{b}})^{2}\right) +{\frac{\varepsilon d}{2}}\int
d^{2}x\,E_{z}^{2}  \notag \\
=& {\frac{\varepsilon d_{w}L^{2}}{2}}\left( (E_{y}^{\text{f}})^{2}+(E_{y}^{%
\text{b}})^{2}\right) +\frac{\varepsilon L^{4}}{24d}(E_{y}^{\text{f}}-E_{y}^{%
\text{b}})^{2},  \label{MaxewBL}
\end{align}%
where $d_{w}$ is the thickness of the layer, and $L$ is the size of the
sample. Note that there is no contribution from the source term due to the
parity, 
\begin{equation}
\int d^{2}xdz\,\varphi (\mathbf{x},z)\rho (\mathbf{x},z)\propto
\int_{-L/2}^{L/2}dy\,y=0.
\end{equation}%
The energy density is given by%
\begin{equation}
\mathcal{H}_{\text{E}}={\frac{\varepsilon d_{w}}{2}}\left( (E_{y}^{\text{f}%
})^{2}+(E_{y}^{\text{b}})^{2}\right) +\frac{\varepsilon L^{2}}{24d}(E_{y}^{%
\text{f}}-E_{y}^{\text{b}})^{2}.
\end{equation}%
The important observation is that the second term diverges in the large
limit of the sample size $L$. The order of the sample size is $L\simeq 1$
mm, while the typical size parameter is $\ell _{B}\simeq d\simeq d_{B}\simeq
10$ nm, as implies $L/\ell _{B}\simeq 10^{6}$. It is a good approximation to
take the limit $L\rightarrow \infty $. We then find%
\begin{equation}
E_{y}^{\text{f}}=E_{y}^{\text{b}}
\end{equation}%
to make the energy density finite.

We rewrite (\ref{HallCurre}) as\beginABC\label{HallCurreE}%
\begin{align}
E_{y}^{\text{f}}=& \frac{2\hbar }{e^{2}\ell _{B}^{2}\rho _{0}}\left[ \frac{%
\mathcal{J}_{x}^{\text{f}}}{1+\sigma _{0}}-\frac{e}{\hbar }(1-\sigma
_{0})J_{s}^{d}\partial _{x}\vartheta \right] , \\
E_{y}^{\text{b}}=& \frac{2\hbar }{e^{2}\ell _{B}^{2}\rho _{0}}\left[ \frac{%
\mathcal{J}_{x}^{\text{b}}}{1-\sigma _{0}}+\frac{e}{\hbar }(1+\sigma
_{0})J_{s}^{d}\partial _{x}\vartheta \right] .
\end{align}%
\endABC The condition $E_{y}^{\text{f}}=E_{y}^{\text{b}}$ requires%
\begin{equation}
E_{y}^{\text{f}}-E_{y}^{\text{b}}=\frac{2\hbar }{e^{2}\ell _{B}^{2}\rho _{0}}%
\left( \frac{\mathcal{J}_{x}^{\text{f}}}{1+\sigma _{0}}-\frac{\mathcal{J}%
_{x}^{\text{b}}}{1-\sigma _{0}}-\frac{2eJ_{s}^{d}}{\hbar }\partial
_{x}\vartheta \right) =0,
\end{equation}%
or%
\begin{equation}
\partial _{x}\vartheta =\frac{\hbar }{2eJ_{s}^{d}}\left( \frac{\mathcal{J}%
_{x}^{\text{f}}}{1+\sigma _{0}}-\frac{\mathcal{J}_{x}^{\text{b}}}{1-\sigma
_{0}}\right) .  \label{PhaseBL}
\end{equation}%
Substituting (\ref{PhaseBL}) into (\ref{HallCurreE}) we obtain%
\begin{equation}
E_{y}^{\text{f}}=E_{y}^{\text{b}}=\frac{\hbar }{e^{2}\ell _{B}^{2}\rho _{0}}%
\left( \mathcal{J}_{x}^{\text{f}}+\mathcal{J}_{x}^{\text{b}}\right) .
\end{equation}%
We conclude that the Hall resistance is given by \beginABC\label{HallResisBL}%
\begin{align}
R_{xy}^{\text{f}}\equiv & {\frac{E_{y}^{\text{f}}}{\mathcal{J}_{x}^{\text{f}}%
}=}\frac{2\pi \hbar }{\nu e^{2}}\left( 1+\frac{\mathcal{J}_{x}^{\text{b}}}{%
\mathcal{J}_{x}^{\text{f}}}\right) , \\
R_{xy}^{\text{b}}\equiv & {\frac{E_{y}^{\text{b}}}{\mathcal{J}_{x}^{\text{b}}%
}=}\frac{2\pi \hbar }{\nu e^{2}}\left( 1+\frac{\mathcal{J}_{x}^{\text{f}}}{%
\mathcal{J}_{x}^{\text{b}}}\right)
\end{align}%
\endABC in each layer. We note that both the diagonal and Hall resistances
are independent of the imbalance parameter $\sigma _{0}$.

\begin{figure}[t]
\includegraphics[width=0.42\textwidth]{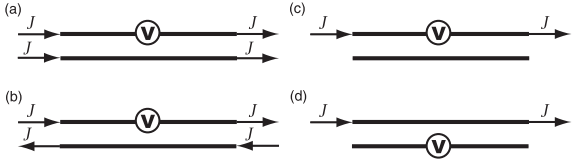}
\caption{ (a) The same amount of current flows on both layers in the same
direction. (b) The same amount of current flows on both layers in the
opposite directions. (c) and (d) The current flows only on the front layer.
In these experiments the diagonal and Hall resistances are measured at one
of the layers indicated by V in a circle. }
\label{FigJosJun}
\end{figure}

\section{Anomalous Bilayer Hall Currents}

\label{SecAnomaCurre}

We apply these formulas to analyze typical bilayer QH currents [Fig.\ref%
{FigJosJun}], and compare the results with the experimental data\cite%
{Kellogg02,Kellogg04,Tutuc04}. Though the experiments were carried out at
the balanced point ($\sigma _{0}=0$), our results are valid also for
imbalanced configurations ($\sigma _{0}\neq 0$).

\subsection{Experiment (a)}

The same amounts of current are fed to the two layers in the experiment [Fig.%
\ref{FigJosJun}(a)]. Since $\mathcal{J}_{x}^{\text{f}}=\mathcal{J}_{x}^{%
\text{b}}$, we obtain from (\ref{PhaseBL}) that%
\begin{equation*}
\vartheta =\text{constant},
\end{equation*}%
and\beginABC%
\begin{equation}
R_{xy}^{\text{f}}\equiv {\frac{E_{y}^{\text{f}}}{\mathcal{J}_{x}^{\text{f}}}=%
}\frac{4\pi \hbar }{\nu e^{2}},\qquad R_{xy}^{\text{b}}\equiv {\frac{E_{y}^{%
\text{b}}}{\mathcal{J}_{x}^{\text{b}}}=}\frac{4\pi \hbar }{\nu e^{2}}.
\end{equation}%
\endABC This is the standard result of the bilayer QH current.\cite%
{Kellogg04,Tutuc04}

\subsection{Counterflow Experiment (b)}

The counterflow experiment [Fig.\ref{FigJosJun}(b)] is most interesting.
Since $\mathcal{J}_{x}^{\text{f}}=-\mathcal{J}_{x}^{\text{b}}$, we obtain
from (\ref{PhaseBL}) that%
\begin{equation}
\vartheta =\frac{\hbar \mathcal{J}_{{x}}^{\text{f}}}{eJ_{s}^{d}}x+\text{%
constant.}
\end{equation}%
and\beginABC%
\begin{equation}
R_{xy}^{\text{f}}\equiv {\frac{E_{y}^{\text{f}}}{\mathcal{J}_{x}^{\text{b}}}=%
}0,\qquad R_{xy}^{\text{b}}\equiv {\frac{E_{y}^{\text{b}}}{\mathcal{J}_{x}^{%
\text{b}}}=}0.
\end{equation}%
\endABC The result is remarkable since it is against the naive picture of
the QH effect. Recall that the essential signal of the QH effect is
considered to be the development of the plateau. The vanishing of the Hall
resistance in the QH regime is a new phenomenon. This anomalous behavior has
been observed experimentally by Kellogg et al.\cite{Kellogg04} and Tutuc et
al.\cite{Tutuc04} at $\nu =1$, as illustrated in Fig.\ref{FigEisenShay}(a).

\subsection{Drag Current (c) \& (d)}

\begin{figure}[t]
\includegraphics[width=0.47\textwidth]{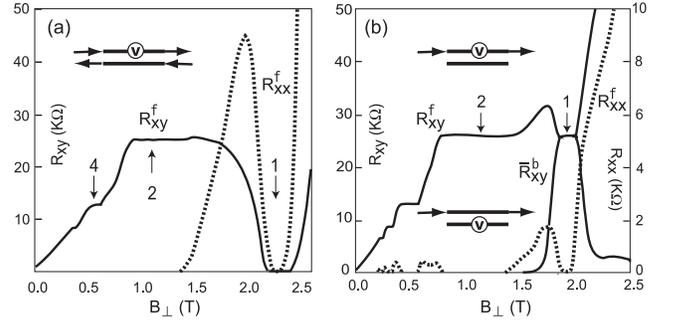}
\caption{ By feeding different currents into the two layers, the Hall
resitance $R_{xy}$ (solid curve) and the longitudinal resistance $R_{xx}$
(dotted curve) are measured on one of the layers (indicated by V in a
circle). (a) In the counterflow experiment the opposite amounts of currents
are fed to the two layers, where $R_{xx}^{\text{f}}=R_{xy}^{\text{f}}=0$
anomalously at $\protect\nu =1$. Data are taken from Ref.\protect\cite%
{Tutuc04}. (b) In the drag experiment the current is fed only to the front
layer, where $R_{xy}^{\text{f}}$ takes anomalously the same value at $%
\protect\nu =1$ and $\protect\nu =2$. It is also remarkable that $R_{xy}^{%
\text{f}}=\bar{R}_{xy}^{\text{b}}$ at $\protect\nu =1$, where $\bar{R}_{xy}^{%
\text{b}}\equiv E_{y}^{\text{b}}/\mathcal{J}_{x}^{\text{f}}$. Data are taken
from Ref.\protect\cite{Kellogg02}.}
\label{FigEisenShay}
\end{figure}

The drag experiment [Figs.\ref{FigJosJun}(c) and (d)] is also very
interesting, where the current is fed only to the front layer. The Hall
resistance is measured in the front layer [Fig.(c)] and also in the back
layer [Figs.(d)]. Since $\mathcal{J}_{x}^{\text{b}}=0$, we obtain from (\ref%
{PhaseBL}) that%
\begin{equation}
\vartheta =\frac{\hbar \mathcal{J}_{x}^{\text{f}}}{2eJ_{s}^{d}}x+\text{%
constant}.
\end{equation}%
In the drag experiment the definition (\ref{HallResisBL}) for $R_{xy}^{\text{%
b}}$ becomes meaningless since $\mathcal{J}_{i}^{\text{b}}=0$. We adopt the
definition%
\begin{equation}
\bar{R}_{xy}^{\text{b}}\equiv {\frac{E_{y}^{\text{b}}}{\mathcal{J}_{x}^{%
\text{f}}}}
\end{equation}%
with the use of $\mathcal{J}_{x}^{\text{f}}$. Then, we find\beginABC%
\begin{equation}
R_{xy}^{\text{f}}\equiv {\frac{E_{y}^{\text{f}}}{\mathcal{J}_{x}^{\text{f}}}=%
}\frac{2\pi \hbar }{\nu e^{2}},\qquad \bar{R}_{xy}^{\text{b}}\equiv {\frac{%
E_{y}^{\text{b}}}{\mathcal{J}_{x}^{\text{f}}}=}\frac{2\pi \hbar }{\nu e^{2}}.
\end{equation}%
\endABC In particular, we have%
\begin{equation}
R_{xy}^{\text{f}}=\bar{R}_{xy}^{\text{b}}={\frac{E_{y}^{\text{f}}}{\mathcal{J%
}_{x}^{\text{f}}}}=\frac{2\pi \hbar }{e^{2}}\qquad \text{at\quad }\nu =1.
\label{DragNU1}
\end{equation}%
On the other hand, if there is no interlayer coherence, the QH current in
the front layer is%
\begin{equation}
\mathcal{J}_{i}^{\text{f}}=\frac{\nu e^{2}}{4\pi \hbar }\varepsilon
_{ij}E_{j}^{\text{f}}
\end{equation}%
at the balance point, and%
\begin{equation}
R_{xy}^{\text{f}}={\frac{E_{y}^{\text{f}}}{\mathcal{J}_{x}^{\text{f}}}}=%
\frac{4\pi \hbar }{\nu e^{2}}.
\end{equation}%
In particular we have%
\begin{equation}
R_{xy}^{\text{f}}={\frac{E_{y}^{\text{f}}}{\mathcal{J}_{x}^{\text{f}}}}=%
\frac{2\pi \hbar }{e^{2}}\qquad \text{at\quad }\nu =2.  \label{DragNU2}
\end{equation}%
It is prominent that from (\ref{DragNU1}) and (\ref{DragNU2}) the Hall
resistance is the same at $\nu =1$ and $\nu =2$. These theoretical results
explain the drag experimental data due to Kellogg et al.\cite{Kellogg02}, as
illustrated in Fig.\ref{FigEisenShay}(b).

\section{Conclusion}

In this paper we have analyzed the dynamics of electrons confined to the
lowest Landau level based on noncommutative geometry. In ordinary physics
the dynamics arises from the kinetic Hamiltonian. In the QH system, however,
it arises from the very nature of noncommutative geometry, that is the W$%
_{\infty }$(N) algebra (\ref{AlgebWpre}) satisfied by the bare density $\hat{%
D}_{\mu \nu }(\mathbf{p})$.

As an application we have derived the formula for the electric current. We
have found that the Coulomb interaction yields quite complicated
contributions through the exchange term to the current. Nevertheless, we
reproduce the standard formula for the Hall current in the monolayer QH
ferromagnet. However, the Hall current contains the phase current in the
bilayer QH ferromagnet. It is a dynamical problem how the phase current
flows. We have shown that it flows in such a way that the Hall current
behaves anomalously as discovered in recent experiments\cite%
{Kellogg02,Kellogg04,Tutuc04}. Furthermore, the anomalous Hall resistance is
unchanged even if the density imbalance is made between the two layers.
These experimental data provide us with another proof of the interlayer
phase coherence spontaneously developed in the bilayer system.

\appendix

\section{SU(2) Effective Hamiltonian}

In this appendix we derive the classical Hamiltonian (\ref{ClassHamilBL})
from the field-theoretical Hamiltonian (\ref{HamilBL}), or $H=H_{\text{C}%
}^{+}+H_{\text{C}}^{-}+H_{\text{T}}+H_{\text{gate}}+H_{\text{E}}$. Each term
is given in terms of the electron density $\rho (\mathbf{x})$ and the
pseudospin density $P_{a}(\mathbf{x})$ as follows,\beginABC\label{BLHamils}%
\begin{align}
H_{\text{C}}^{+}& ={\frac{1}{2}}\int \!d^{2}xd^{2}y\;V^{+}(\mathbf{x}-%
\mathbf{y})\rho (\mathbf{x})\rho (\mathbf{y}),  \label{BLCouloP} \\
H_{\text{C}}^{-}& =2\int \!d^{2}xd^{2}y\;V^{-}(\mathbf{x}-\mathbf{y})P_{z}(%
\mathbf{x})P_{z}(\mathbf{y}),  \label{BLCouloM} \\
H_{\text{T}}& =-\Delta _{\text{SAS}}\int \!d^{2}x\;P_{x}(\mathbf{x}), \\
H_{\text{gate}}& =-\Delta _{\text{bias}}\int \!d^{2}x\;P_{z}(\mathbf{x}),
\label{BiasTerm} \\
H_{\text{E}}& =-e\int \!d^{2}x\,\left[ \varphi ^{\text{f}}(\mathbf{x})\rho ^{%
\text{f}}(\mathbf{x})+\varphi ^{\text{b}}(\mathbf{x})\rho ^{\text{b}}(%
\mathbf{x})\right] ,
\end{align}%
\endABC where $\Delta _{\text{bias}}$ is the bias parameter to take care of
the charge imbalance made by the gate voltages in (\ref{BiasTerm}). The
Coulomb potentials are%
\begin{equation}
V^{\pm }(\boldsymbol{x})=\frac{e^{2}}{8\pi \epsilon }\bigg(\frac{1}{|%
\boldsymbol{x}|}\pm \frac{1}{\sqrt{|\boldsymbol{x}|^{2}+d^{2}}}\bigg).
\label{CouloPM}
\end{equation}%
We take the expectation value by the Fock state (\ref{SkyrmFormuC}). The
Coulomb energy is decomposed into the direct and exchange energies\cite%
{Ezawa05D},\beginABC%
\begin{align}
& \langle H_{\text{C}}^{+}\rangle =\pi \int d^{2}pV_{\text{D}}^{+}(%
\boldsymbol{p})\hat{\varrho}(-\boldsymbol{p})\hat{\varrho}(\boldsymbol{p}) 
\notag \\
& \hspace{1cm}-\pi \int d^{2}pV_{\text{X}}^{+}(\boldsymbol{p})\bigg[\mathcal{%
P}_{z}(-\boldsymbol{p})\mathcal{P}_{z}(\boldsymbol{p})+\frac{1}{4}\varrho (-%
\boldsymbol{p})\varrho (\boldsymbol{p})\bigg]  \notag \\
& \hspace{1.4cm}-\pi \sum_{a=x,y}\int d^{2}pV_{\text{X}}^{+}(\boldsymbol{p})%
\mathcal{P}_{a}(-\boldsymbol{p})\mathcal{P}_{a}(\boldsymbol{p}), \\
& \langle H_{\text{C}}^{-}\rangle =4\pi \int d^{2}pV_{\text{D}}^{-}(%
\boldsymbol{p})\mathcal{P}_{z}(-\boldsymbol{p})\mathcal{P}_{z}(\boldsymbol{p}%
)  \notag \\
& \hspace{1cm}-\pi \int d^{2}pV_{\text{X}}^{-}(\boldsymbol{p})\bigg[\mathcal{%
P}_{z}(-\boldsymbol{p})\mathcal{P}_{z}(\boldsymbol{p})+\frac{1}{4}\varrho (-%
\boldsymbol{p})\varrho (\boldsymbol{p})\bigg]  \notag \\
& \hspace{1.4cm}+\pi \sum_{a=x,y}\int d^{2}pV_{\text{X}}^{-}(\boldsymbol{p})%
\mathcal{P}_{a}(-\boldsymbol{p})\mathcal{P}_{a}(\boldsymbol{p}).
\end{align}%
\endABC All other terms are simply converted into the corresponding
classical terms,%
\begin{align}
& \langle H_{\text{T}}\rangle =-2\pi \Delta _{\text{SAS}}\mathcal{P}_{x}(0),
\\
& \langle H_{\text{gate}}\rangle =-2\pi \Delta _{\mathsf{bias}}\emph{P}%
_{z}(0), \\
& \langle H_{\text{E}}\rangle =-e\int \!d^{2}q\,e^{-\mathbf{q}^{2}\ell
_{B}^{2}/4}\left[ \varphi ^{\text{f}}(-\mathbf{q})\varrho ^{\text{f}}(%
\mathbf{q})+\varphi ^{\text{b}}(-\mathbf{q})\varrho ^{\text{b}}(\mathbf{q})%
\right] .
\end{align}%
We now analyze the ground-state condition. We substitute%
\begin{equation}
\varrho (\mathbf{x})=\rho _{0},\quad \emph{P}_{a}(\mathbf{x})={\frac{1}{2}%
\rho }_{0}(\sqrt{1-\sigma _{0}},0,\sigma _{0})  \label{GrounState}
\end{equation}%
into the total energy $\mathcal{H}\equiv \langle H\rangle $, and minimize it
with respect to $\sigma _{0}$. As a result we obtain the condition%
\begin{equation}
\Delta _{\text{bias}}=\frac{\sigma _{0}}{\sqrt{1-\sigma _{0}^{2}}}\Delta _{%
\text{\textsf{SAS}}}+\sigma _{0}\epsilon _{\text{cap}}  \label{CondiGrounBL}
\end{equation}%
with (\ref{CapacParam}). Eliminating the parameter $\Delta _{\text{bias}}$
from the total classical Hamiltonian 
\begin{equation}
\mathcal{H}=\langle H_{\text{C}}^{+}\rangle +\langle H_{\text{C}}^{-}\rangle
+\langle H_{\text{T}}\rangle +\langle H_{\text{gate}}\rangle +\langle H_{%
\text{E}}\rangle ,
\end{equation}%
we can rearrange it into%
\begin{equation}
\mathcal{H}=\mathcal{H}_{\text{D}}+\mathcal{H}_{\text{X}}+\mathcal{H}_{\text{%
T}}+\mathcal{H}_{\text{bias}}+\mathcal{H}_{\text{E}},  \label{EffecHamilBL}
\end{equation}%
where various terms are given by (\ref{ClassHamilBL}) in text.

We proceed to analyze a small fluctuation of the pseudospin field $\emph{P}%
_{a}(\mathbf{x})$ around the ground state (\ref{GrounState}), which
describes the pseudospin wave. We parametrize%
\begin{equation}
\emph{P}_{a}(\mathbf{x})=\frac{\rho _{0}}{2}\mathbf{n}^{\dag }(\mathbf{x}%
)\pi _{a}\mathbf{n}(\mathbf{x}),  \label{PpinField}
\end{equation}%
where $\pi _{a}$ is the Pauli matrix, and%
\begin{align}
\mathbf{n}(\mathbf{x})& =\frac{1}{\sqrt{2}}\left( 
\begin{array}{cc}
\sqrt{1+\sigma _{0}} & \sqrt{1-\sigma _{0}} \\ 
\sqrt{1-\sigma _{0}} & -\sqrt{1+\sigma _{0}}%
\end{array}%
\right) \left( 
\begin{array}{c}
\sqrt{1-|\eta (\mathbf{x})|^{2}} \\ 
\eta (\mathbf{x})%
\end{array}%
\right)  \notag \\
& \simeq \frac{1}{\sqrt{2}}\left( 
\begin{array}{cc}
\sqrt{1+\sigma _{0}} & \sqrt{1-\sigma _{0}} \\ 
\sqrt{1-\sigma _{0}} & -\sqrt{1+\sigma _{0}}%
\end{array}%
\right) \left( 
\begin{array}{c}
1 \\ 
\eta (\mathbf{x})%
\end{array}%
\right)  \label{FieldCP}
\end{align}%
in the linear approximation, with%
\begin{equation}
\eta (\mathbf{x})=\frac{\sigma (\mathbf{x})+i\vartheta (\mathbf{x})}{2}%
,\quad \eta ^{\dagger }(\mathbf{x})=\frac{\sigma (\mathbf{x})-i\vartheta (%
\mathbf{x})}{2}.  \label{EtaFieldBL4}
\end{equation}%
The pseudospin field (\ref{PpinField}) is reduced to the ground state (\ref%
{GrounState}) for $\eta (\mathbf{x})=0$. Substituting the pseudospin field (%
\ref{PpinField}) into (\ref{EffecHamilBL}) together with this
parametrization, we find%
\begin{align}
\mathcal{H}=& \frac{(1-\sigma _{0}^{2})J_{s}+\sigma _{0}^{2}J_{s}^{d}}{2}%
(\partial _{k}\sigma )^{2}  \notag \\
& +\frac{\rho _{0}}{4}\left[ \epsilon _{\text{cap}}(1-\sigma _{0}^{2})+\frac{%
\Delta _{\text{SAS}}}{\sqrt{1-\sigma _{0}^{2}}}\right] \sigma ^{2}  \notag \\
& +\frac{1}{2}J_{s}^{d}(\partial _{k}\vartheta )^{2}+\frac{\rho _{0}}{4}%
\frac{\Delta _{\text{SAS}}}{\sqrt{1-\sigma _{0}^{2}}}\vartheta ^{2}
\label{PpinWave}
\end{align}%
up to the second order in $\sigma (\mathbf{x})$ and $\vartheta (\mathbf{x})$%
. Their coherence lengths are%
\begin{align}
\xi _{\vartheta }& =2\ell _{B}\sqrt{{\frac{\pi \sqrt{1-\sigma _{0}^{2}}%
J_{s}^{d}}{\Delta _{\text{SAS}}}}},  \notag \\
\xi _{\sigma }& =2\ell _{B}\sqrt{{\frac{\pi \left[ (1-\sigma
_{0}^{2})J_{s}+\sigma _{0}^{2}J_{s}^{d}\right] }{\epsilon _{\text{cap}%
}(1-\sigma _{0}^{2})+\Delta _{\text{SAS}}/\sqrt{1-\sigma _{0}^{2}}}}}.
\label{CoherLengtPpin}
\end{align}%
It is remarkable that the coherent length $\xi _{\vartheta }$ becomes
infinitely large as $\Delta _{\text{SAS}}\rightarrow 0$, though $\xi
_{\sigma }$ remains finite due to the capacitance-energy parameter $\epsilon
_{\text{cap}}$. It follows from (\ref{CondiGrounBL}) that the condition $%
\Delta _{\text{bias}}<\epsilon _{\text{cap}}$ is necessary to obtain the
gapless mode.

\section{SU(4) Effective Hamiltonian}

We have ignored the spin degree of freedom when we have analyzed the Hall
currents in the bilayer QH system in Section \ref{SecBLQHcurre}. In this
appendix, including all components of the SU(4) QH system, we derive the
effective Hamiltonian and justify this simplification.

There are 15 isospin components,%
\begin{eqnarray}
S_{a} &=&{\frac{1}{2}}\Psi ^{\dagger }\sigma _{a}\Psi (\mathbf{x}),\quad
P_{a}={\frac{1}{2}}\Psi ^{\dagger }\pi _{a}\Psi ,  \notag \\
R_{ab} &=&\frac{1}{2}\Psi ^{\dagger }\sigma _{a}\pi _{b}\Psi .
\end{eqnarray}%
We denote their classical fields as $\mathcal{S}_{a}$, $\mathcal{P}_{a}$ and 
$\mathcal{R}_{ab}$ as in (\ref{ClassS}) and (\ref{ClassP}). The classical
Hamiltonian has already been derived\cite{Ezawa05D}, where the Coulomb
energy density is%
\begin{align}
\langle H_{\text{C}}^{\text{cl}}\rangle =& \pi V_{\text{D}}^{+}(\mathbf{p}%
)\varrho (-\mathbf{p})\varrho (\mathbf{p})+4\pi V_{\text{D}}^{-}(\mathbf{p})%
\mathcal{P}_{z}(-\mathbf{p})\mathcal{P}_{z}(\mathbf{p})  \notag \\
& -\frac{\pi }{2}V_{\text{X}}^{d}(\mathbf{p})[\mathcal{S}_{a}(-\mathbf{p})%
\mathcal{S}_{a}(\mathbf{p})+\mathcal{P}_{a}(-\mathbf{p})\mathcal{P}_{a}(%
\mathbf{p})  \notag \\
& \hspace*{2.8cm}+\mathcal{R}_{ab}(-\mathbf{p})\mathcal{R}_{ab}(\mathbf{p})]
\notag \\
& -\pi V_{\text{X}}^{-}(\mathbf{p})[\mathcal{S}_{a}(-\mathbf{p})\mathcal{S}%
_{a}(\mathbf{p})+\mathcal{P}_{z}(-\mathbf{p})\mathcal{P}_{z}(\mathbf{p}) 
\notag \\
& \hspace*{2.8cm}+\mathcal{R}_{az}(-\mathbf{p})\mathcal{R}_{az}(\mathbf{p})]
\notag \\
& -\frac{\pi }{8}V_{\text{X}}(\mathbf{p})\varrho (-\mathbf{p})\varrho (%
\mathbf{p}).
\end{align}%
The Zeeman term, the tunneling term and the bias terms are given by (\ref%
{HamilClassZ}), (\ref{HamilClassT}) and (\ref{HamilClassB}), respectively.

The ground state is given by%
\begin{align}
\mathcal{S}_{a}^{\text{g}}& =\frac{1}{2}\delta _{az},\qquad \mathcal{P}_{a}^{%
\text{g}}=\frac{1}{2}(\sqrt{1-\sigma _{0}^{2}}\delta _{ax}+\sigma _{0}\delta
_{az}),  \notag \\
\mathcal{R}_{ab}^{\text{g}}& =\frac{1}{2}\delta _{az}(\sqrt{1-\sigma _{0}^{2}%
}\delta _{bx}+\sigma _{0}\delta _{bz})  \label{GrounSPin}
\end{align}%
in the imbalanced configuration at $\sigma _{0}$. We analyze a small
fluctuation of the isospin field around the ground state. We may parametrize%
\begin{equation}
\emph{S}_{a}=\frac{\rho _{0}}{2}\mathbf{n}^{\dag }\sigma _{a}\mathbf{n}%
,\quad \emph{P}_{a}=\frac{\rho _{0}}{2}\mathbf{n}^{\dag }\pi _{a}\mathbf{n}%
,\quad \emph{R}_{ab}=\frac{\rho _{0}}{2}\mathbf{n}^{\dag }\sigma _{a}\pi _{b}%
\mathbf{n,}  \label{SPR}
\end{equation}%
where%
\begin{equation}
\mathbf{n}(\mathbf{x})=\frac{1}{\sqrt{2}}\left( 
\begin{array}{cccc}
f_{+} & 0 & f_{-} & 0 \\ 
0 & f_{+} & 0 & f_{-} \\ 
f_{-} & 0 & -f_{+} & 0 \\ 
0 & f_{-} & 0 & -f_{+}%
\end{array}%
\right) \left( 
\begin{array}{c}
1 \\ 
\eta _{\text{s}}(\mathbf{x}) \\ 
\eta _{\text{p}}(\mathbf{x}) \\ 
\eta _{\text{r}}(\mathbf{x})%
\end{array}%
\right)  \label{CP4}
\end{equation}%
in the linear approximation as in (\ref{FieldCP}), with $f_{\pm }=\sqrt{1\pm
\sigma _{0}}$. The phase field $\vartheta _{i}(\mathbf{x})$ and the
imbalance field $\sigma _{i}(\mathbf{x})$ are introduced as in (\ref%
{EtaFieldBL4}) for each component, $i=$s, p, r.

The effective Hamiltonian is derived by substituting (\ref{SPR}) together
with (\ref{CP4}) into the classical Hamiltonians and making the derivative
expansions. It is found that the pseudospin mode $\eta _{\text{p}}(\mathbf{x}%
)$ is described precisely by the Hamiltonian (\ref{PpinWave}). To study the
other two modes, we change the variables as%
\begin{align}
\eta _{s}& =\sqrt{\frac{1+\sigma _{0}}{2}}\eta _{1}+\sqrt{\frac{1-\sigma _{0}%
}{2}}\eta _{2},  \notag \\
\eta _{r}& =\sqrt{\frac{1-\sigma _{0}}{2}}\eta _{1}-\sqrt{\frac{1+\sigma _{0}%
}{2}}\eta _{2}.
\end{align}%
The effective Hamiltonian reads%
\begin{align}
\mathcal{H}_{\text{mix}}=& \frac{J_{s}^{+}+\sigma _{0}J_{s}^{-}}{2}%
[(\partial _{k}\sigma _{1})^{2}+(\partial _{k}\vartheta _{1})^{2}]  \notag \\
& +\frac{\rho _{0}}{4}\left( \Delta _{\text{Z}}+\frac{1}{2}\Delta _{\text{SAS%
}}\sqrt{\frac{1-\sigma _{0}}{1+\sigma _{0}}}\right) [\sigma
_{1}^{2}+\vartheta _{1}^{2}]  \notag \\
& +\frac{J_{s}^{+}-\sigma _{0}J_{s}^{-}}{2}[(\partial _{k}\sigma
_{2})^{2}+(\partial _{k}\vartheta _{2})^{2}]  \notag \\
& +\frac{\rho _{0}}{4}\left( \Delta _{\text{Z}}+\frac{1}{2}\Delta _{\text{SAS%
}}\sqrt{\frac{1+\sigma _{0}}{1-\sigma _{0}}}\right) [\sigma
_{2}^{2}+\vartheta _{2}^{2}]  \notag \\
& +\frac{\rho _{0}}{4}\Delta _{\text{SAS}}(\sigma _{1}\sigma _{2}+\vartheta
_{1}\vartheta _{2}),
\end{align}%
where%
\begin{equation}
J_{s}^{\pm }=\frac{1}{2}\left( J_{s}\pm J_{s}^{d}\right)  \label{IntroStiffC}
\end{equation}%
with (\ref{StiffPM}) and (\ref{StiffParam}). The two modes $\eta _{1}$ and $%
\eta _{2}$ are coupled in general, but decoupled for $\Delta _{\text{SAS}}=0$%
. There exist no gapless modes in this Hamiltonian provided $\Delta _{\text{Z%
}}\neq 0$.

In conclusion, when $\Delta _{\text{SAS}}=0$, there is only one gapless
mode, which is the interlayer phase field $\vartheta _{\text{p}}$. This
justifies that we have neglected all dynamical fields except for the field $%
\vartheta _{\text{p}}$ to analyze the currents in Section \ref{SecAnomaCurre}%
.


\bigskip

\begin{thebibliography}{99}
\bibitem{EzawaBook} {\ Z.F. Ezawa, \textit{Quantum Hall Effects:
Field-Theoretical Approach and Related Topics} (World Scientific, 2000). }

\bibitem{BookDasSarma} S. Das Sarma and A. Pinczuk (eds), \textit{%
Perspectives in Quantum Hall Effects} (Wiley, 1997).

\bibitem{Kellogg04} M. Kellogg, J.P. Eisenstein, L.N. Pfeiffer and K.W.
West, Phys. Rev. Lett. \textbf{93} (2004) 036801.

\bibitem{Tutuc04} E. Tutuc, M. Shayegan and D.A. Huse, Phys. Rev. Lett. 
\textbf{93} (2004) 036802.

\bibitem{Kellogg02} M. Kellogg, I.B. Spielman, J.P. Eisenstein, L.N.
Pfeiffer and K.W. West, Phys. Rev. Lett. \textbf{88} (2002) 126804.

\bibitem{Eisenstein04} J.P. Eisenstein and A.H. MacDonald, Nature \textbf{432%
} (2004) 691.

\bibitem{Girvin86B} {\ S.M. Girvin, A.H. MacDonald and P.M. Platzman, Phys.
Rev. B \textbf{33} (1986) 2481.}

\bibitem{Girvin84B} S.M. Girvin and T. Jach, Phys. Rev. B \textbf{29} (1984)
5617.

\bibitem{Girvin85L} S.M. Girvin, A.H. MacDonald and P.M. Platzman, Phys.
Rev. Lett. \textbf{54} (1985) 581.

\bibitem{Stone93IJMPB} M. Stone and J. Martinez, Int. J. Mod. Phys. B 
\textbf{7} (1993) 4389.

\bibitem{Ezawa05D} Z.F. Ezawa and G. Tsitsishvili, Phys. Rev. D \textbf{72}
(2005) 85002; {G. Tsitsishvili and Z.F. Ezawa, Phys. Rev. B \textbf{72}
(2005) 115306. }

\bibitem{BookConnes} A. Connes, Noncommutative Geometry (Academic Press,
1994).

\bibitem{Ezawa92IJMPB} {\ Z.F. Ezawa and A. Iwazaki, Int. J. Mod. Phys. B 
\textbf{6} (1992) 3205. }

\bibitem{Moon95B} K. Moon, H. Mori, K. Yang, S.M. Girvin, A.H. MacDonald, L.
Zheng, D. Yoshioka and S-C. Zhang, Phys. Rev. B \textbf{51} (1995) 5138.

\bibitem{EzawaX03B} {\ Z.F. Ezawa, G. Tsitsishvili and K. Hasebe, Phys. Rev.
B \textbf{67} (2003) 125314.}
\end{thebibliography}
\end{document}